\documentclass{article}
\usepackage{arxiv}
\usepackage{hyperref}       
\usepackage{url}            
\usepackage{booktabs}       
\usepackage{amsfonts}       
\usepackage{nicefrac}       
\usepackage{microtype}      
\usepackage{lipsum}
\usepackage{graphicx}
\usepackage{algorithm}
\usepackage{graphicx}
\usepackage{epstopdf}
\usepackage{soul}
\usepackage[noend]{algpseudocode}
\usepackage{xcolor}
\usepackage{amsmath}
\def\BState{\State\hskip-\ALG@thistlm}
\makeatother
\usepackage[acronym,nonumberlist]{glossaries}

\makeglossaries 

\title{KupenStack: Kubernetes based Cloud Native OpenStack
}

\author{
 Parth Yadav \\
  Ramanujan College,\\ 
  University of Delhi\\
  \texttt{parthyadav3105@gmail.com}\\
  \And
  Vipin Kumar Rathi \\
  Ramanujan College,\\ University of Delhi\\
  \texttt{vipin.rathi@ramanujan.du.ac.in}\\
}

\begin{document}

\maketitle

\begin{abstract}
OpenStack is  an open-source private cloud used to run VMs and its related cloud services. OpenStack deployment, management, and upgradation require lots of efforts and manual troubleshooting. Also, workloads and services offered by OpenStack cannot self-heal itself on failures. We present KupenStack, a Cloud-Native OpenStack as Code model built on top of Kubernetes stack as Custom Resources. KupenStack is a controller that interacts between Kubernetes and OpenStack and automates complex operations like scaling, LCM, zero-downtime, self-healing, version upgrades, configuration management, and offers OpenStack as a service through code. KupenStack builds cloud-native values like immutable infrastructure, declarative APIs for OpenStack without changing any OpenStack code. If a VM workload goes down for some reason, then KupenStack handles it and automatically spins up a new instance. KupenStack uses OpenStack on Kubernetes deployment for lifecycle management of OpenStack.       
\end{abstract}



\section{Introduction}

OpenStack is a private cloud and manages end-to-end resources needed in a private cloud from Bare metal to data plane to control plane. OpenStack was mainly used to manage virtual machines, but it is also used to manage Containers over time. There are various implementations available for deploying OpenStack i.e., OpenStack On Kubernetes(OOK),  OpenStack on OpenStack(OOO); among various deployments, each deployment has its own pros and cons. Among all, OOK has the least overhead and better management. Airship is one of the projects which uses OOK. Airship automates cloud provisioning and management by establishing Under Cloud Platform(UCP) leveraging Kubernetes. Airship also uses OpenStack-Helm for deployment activities. For interoperability and better adoption, a new project was launched in OpenStack named as OpenStack-Helm.  OpenStack-Helm provides a collection of Helm charts that simply, resiliently, and flexibly deploy OpenStack and related services on Kubernetes. OpenStack Helm helps in making OpenStack containerized means each of its services like Nova, Keystone; Neutron is running from a container. Currently, many Container Orchestration Engines are available like Docker Swarm, Kubernetes, Apache Mesos, etc. For managing containers, CNCF launched Kubernetes. It is an orchestrator that manages Life Cycle Management of Containers. Kubernetes is now a de facto for Multi-cloud. Multi-cloud means workload can be switched between different public clouds, e.g., AWS, Azure, GCP.
Kubernetes also plays a key role in Hybrid Cloud. In a hybrid cloud, workloads can be switched between any private cloud and a different public cloud. Kubernetes installs containerized OpenStack components with the help of OpenStack-Helm in a declarative way.\\
OpenStack Deployment, Management and Upgradation (DMU) is a tedious task and requires many skills. To overcome this and provide abstraction, we propose KupenStack, which automates the DMU and provides a seamless experience for developers. KupenStack can also be used in both hybrid or multi-cloud like Kubernetes. KupenStack can provide consistent, repeatable, easy, scalable, customizable OpenStack deployments.

\section{Cloud-Native OpenStack as-code on Kubernetes}
KupenStack is an OOK controller.  KupenStack architecture contains controller patterns with the automated implementation of complex operations. KupenStack provides OpenStack services(like Nova, Neutron) and resources(like VM, Subnet ) as custom resources. Instead of building  Cloud-Native OpenStack from scratch, KupenStack proposes to reuse Kubernetes control plane to make OpenStack as Cloud Native. This section starts with an introduction to the KupenStack architecture and design principles. Then covers some observations on mappings between OpenStack and Kubernetes concepts. Finally, giving a reference implementation example of the KupenStack model.
\subsection{Architecture}
OpenStack internally has a distributed service architecture, i.e., loosely coupled services communicate with each other using REST APIs, RPC and messaging queues. This decoupled service architecture of OpenStack gives benefits which include (a) the ability to customize and cherry-pick components to install (b) freedom to scale individual components (c) interoperability to integrate components from multiple vendors. The KupenStack architecture promises to keep these benefits safe by not changing OpenStack implementation in anyways. KupenStack architecture proposes to deploy OpenStack components as containers on top of Kubernetes. Fundamental reasons for KupenStack to choose containers are:

\begin{itemize}
\item \textbf{KupenStack can operate on multi-vendor OpenStack deployments}. KupenStack deploys the OpenStack services as containers and communicates to them with standard OpenStack APIs. This makes individual service images interchangeable with multiple vendor implementations.
\item \textbf{KupenStack can ensure OpenStack deployments that are highly portable}. Container images can run on any Linux kernel-based operating system without any dependency failures. KupenStack components themselves will run as a containerized workload on Kubernetes; this builds trust for high portability.

\item \textbf{KupenStack can achieve OpenStack deployments that are robust and fault-tolerant}. Container images follow the Image Immutability Principle(IIP). KupenStack promises to follow the immutable infrastructure paradigms \cite{WhatIsIm18:online} for configuration and version changes of OpenStack.

\item \textbf{KupenStack can achieve Cloud Native Infrastructure operations}. OOK deployment is possible by using containers. An OOK deployment provides scalability, resiliency, and automation. With OOK deployments, KupenStack can easily build operations for OpenStack like zero-touch provisioning, scaling, lcm, zero-downtime, self-healing, version upgrades, configuration management, and many more. These operations themselves can be cloud-native by an implementation.
\end{itemize}

\begin{figure}[ht]
  \centering
  \includegraphics[width=7in, height=4in]{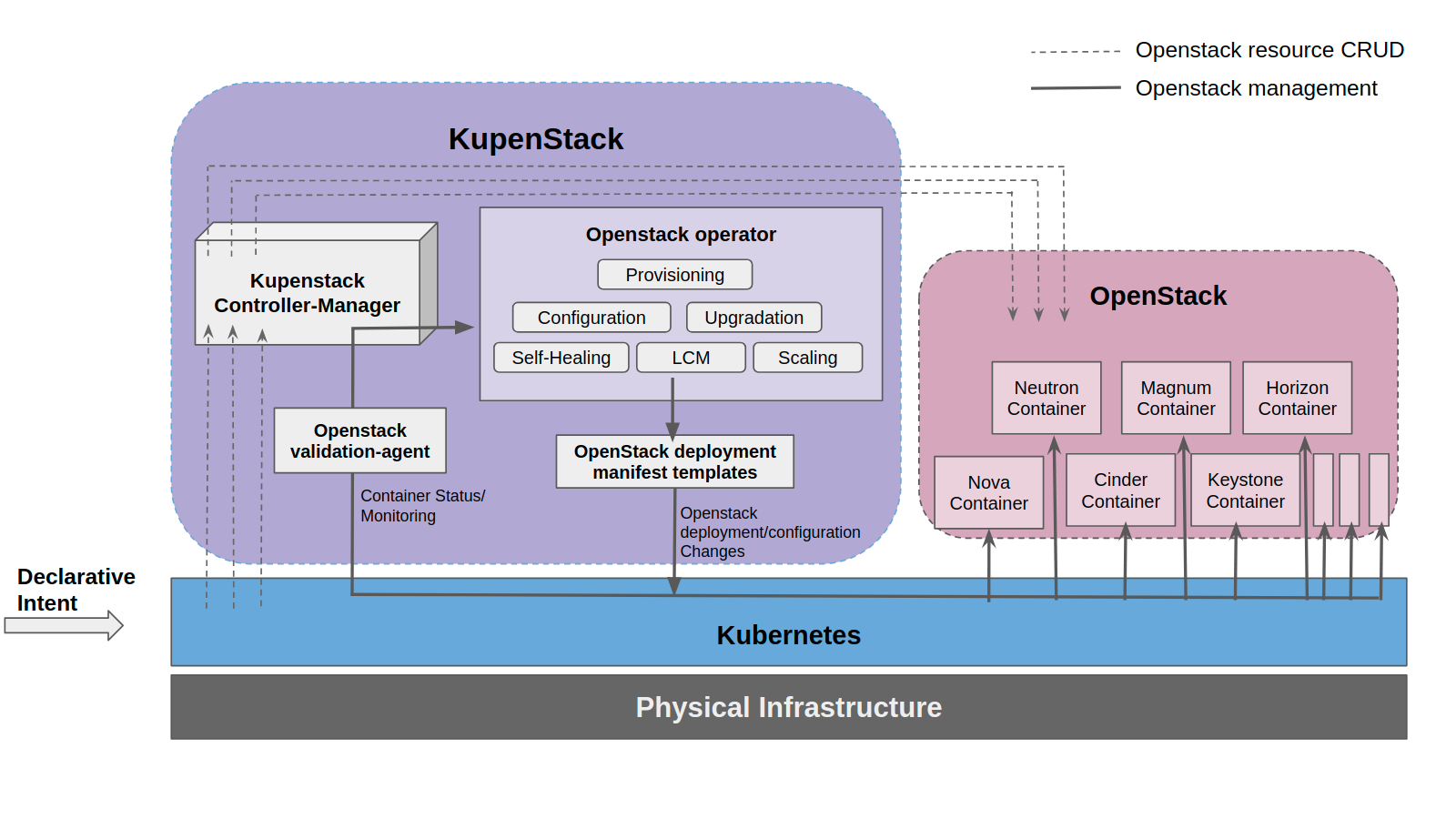}\\
  \caption{KupenStack workflow}\label{kw}
\end{figure}
\begin{figure}[!ht]
  \centering
  \includegraphics[width=5in, height=3in]{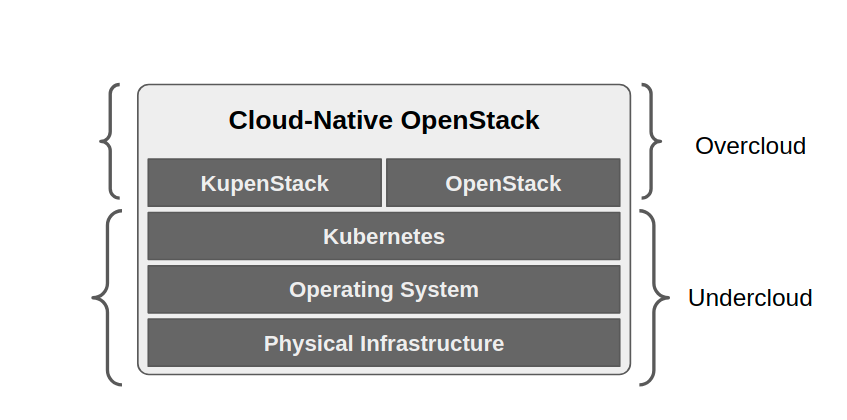}\\
  \caption{ Cloud-Native OpenStack}\label{CNO}
\end{figure}
KupenStack extends Kubernetes using controllers and custom resources to build a cloud-native layer interacting between OpenStack and Kubernetes.
KupenStack follows two principles:

\begin{enumerate}
\item Should not change anything in OpenStack(i.e., Compatibility with any certified OpenStack)
\item Should not change anything in Kubernetes(i.e., Compatibility with any certified Kubernetes)
\end{enumerate}

The degree of cloud-native behavior of custom resources depends on KupenStack implementation.
Fig.\ref{kw} shows working of KupenStack architecture. The purple box represents KupenStack while blue and pink indicate Kubernetes and OpenStack respectively. Two flows are described by Dark arrow and Dotted arrow.
\begin{itemize}
\item Dark Arrow: The user tells what OpenStack components need to be installed and some configurations changes and version changes as declarative intent using custom resources. KupenStack-Controller-Manager notices the required changes and it queues corresponding controllers to take actions. The controller intelligently decides what action is required and calls the OpenStack-Operator to take those actions. The OpenStack-Operator builds Kubernetes manifest files for OpenStack container deployment using some existing templates. Applies those manifest to Kubernetes and then required OpenStack deployment gets spun up on top of Kubernetes. Now, suppose due to some reason some OpenStack service starts failing, then the OpenStack validation agent who is consistently monitoring these services, collecting logs, metrics notifies the required controller that what has failed and other details, logs. Controllers then decide what needs to be done, and if action is required then make a call to the OpenStack operator to make changes.

\item Dotted arrow: The user tells what OpenStack resources are needed to be created, let's say a Virtual Machine as a declarative intent using custom resources. KupenStack-Controller-Manager notices the required changes and it queues corresponding controllers to create a Virtual Machine. The controller makes a REST request to Nova service to create a Virtual Machine with the required spec. Nova creates Virtual Machine and returns an ID. Now, controllers keep watching the status of Virtual Machine for failures and keep updating the status on the custom resource. If for some reason Virtual Machines fails after some time, KupenStack self-heals it or else notifies the user of the cause of the failure.
\end{itemize}
The deployment topology of Cloud-Native OpenStack is shown in Fig.\ref{CNO}\\ \\

\begin{figure}[h]
  \centering
  \includegraphics[width=6in, height=2in]{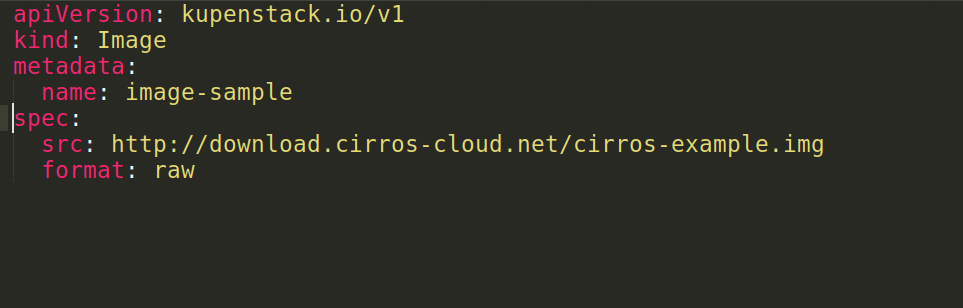}\\
  \caption{ Manifest}\label{ys}
\end{figure}

The above-described architecture gives us the ''Cloud-Native OpenStack as-Code'' pattern. Here ''as-Code'' implies that all APIs are machine-readable definition files, i.e. for using and operating this implementation of OpenStack we need minimum manual operations. We only declare the required state and KupenStack does the heavy-lifting work for managing OpenStack. In practice this would look like following YAML for provisioning virtual machine image in a glance as shown in Fig. \ref{ys}

\subsection{Mapping between OpenStack and Kubernetes}
\begin{figure}[h]
  \centering
  \includegraphics[width=6in, height=2in]{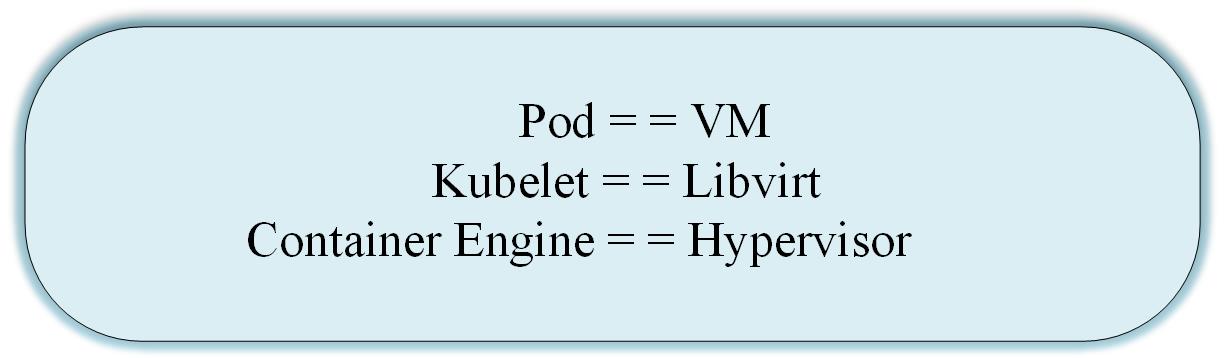}\\
  \caption{Workload Concepts Mapping}\label{ma}
\end{figure}

OpenStack and Kubernetes are similar in some ways, primarily both orchestrate workloads. Discussing all similarities is out of the scope of this paper. However, this section discusses some observations on the mapping between OpenStack and Kubernetes concepts to build some high-level understandings. A specific implementation of KupenStack can follow any approach while mapping OpenStack to Kubernetes through custom resources. KupenStack is a cloud-native layer of interaction between OpenStack and Kubernetes, so it can map similar concepts and provide abstraction. This paper doesn’t mandate that KupenStack should have a particular implementation for mapping resources. Evaluating the best possible mapping of OpenStack resources to Kubernetes custom resources in KupenStack can be further explored.\\\\
Kubernetes cluster contains multiple nodes (bare metal or VM); a node can be either control-plane (Master node) or compute nodes (Worker node) or both.  OpenStack also follows the same kind of architecture as shown in Fig.\ref{m}\\\\
Pods are the primary workload in Kubernetes and were initially designed keeping Virtual Machines in mind. Hence, Pod is a group of containers sharing network namespace, have initContainers, etc. When pods are scheduled to run on a node, Kubelet agent  (on that node) talks to the container engine to create containers. In OpenStack when a Virtual machine is scheduled to run on a node, libvirt agent (on that node) talks to the hypervisor to create a Virtual Machine as shown in Fig. \ref{ma}\\

Namespace logically separates resources in a Kubernetes cluster. Similarly, OpenStack has Projects/Tenants to isolate resources. Some resources in Kubernetes are cluster scoped (like Persistent Volume). In OpenStack,  resources can be used in two ways, i.e., Project-scoped or shared( Cross-project visibility) for example, Network resource in OpenStack can be configured to be shared across Projects or be specific to Project. Hence Namespace in Kubernetes can be mapped to Projects in OpenStack.\\
Domains are the highest level of logical isolation in OpenStack.  Domains are organization-level isolation owning a collection of users, groups, and projects. All resources created in a Domain are owned by that Domain and have no visibility in other Domains. Kubernetes does not have any such implementation which isolates the control plane itself.\\
Both Kubernetes and OpenStack have a Role-Based Access Control (RBAC) to define authorization and access on Namespaces and Projects, respectively, but the implementation of RBAC in both differ slightly. Every OpenStack Project decides which User or User Group should have access to its resources using Roles. Roles have an RBAC policy that defines rules to enforce on each OpenStack API endpoint. A rule defines whether an API endpoint can be accessed by an admin, user, or both.
For RBAC in Kubernetes,  Roles and ClusterRoles are applied on Users or Groups using RoleBindings and ClusterRoleBindings. Each Kubernetes Role and ClusterRole definition can define multiple rules, where each rule defines access for resource types and verbs(actions) allowed on them. Kubernetes and OpenStack both have Roles, but there cannot be any clear mapping between them in terms of usage.

For authentication, both Kubernetes and OpenStack have a notion of Users and User Groups, but both have entirely different authentication systems. However, some work is available on using Keystone users to authenticate Kubernetes \cite{Authenti87:online} \cite{cloudpro53:online} \cite{Strength41:online} as this is a desirable feature for Kubernetes on OpenStack(KOO) for open-source projects like OpenStack Magnum. As of this writing, we have no reference to using Kubernetes users(kubeconfig certificates) to authenticate to OpenStack. While both Kubernetes and OpenStack allow authentication from external identity providers like LDAP.\\\\
OpenStack has hierarchical concepts like Regions, Host Aggregates and Availability Zones to provide logical partitioning within the cluster nodes for failure domains or to meet special hardware needs like sriov, GPU, etc. A Region means complete OpenStack deployment level isolation. It provides separate API endpoints for all services and resource(like Subnet, VM) pools. However, Regions share Keystone and Horizon services. Looking inside a Region, we can have multiple Availability Zones, which in theory are for logical partitioning. In practice, the exact implementation of Availability Zone depends on respective OpenStack services like Nova, Neutron, etc. Availability Zones is one of the most ambiguous topics for beginners. For this section, let us look at Availability Zones for two services, namely Nova and Cinder, to understand this problem.\\\\
In Nova, every Availability Zones is Host Aggregates but not vice versa. Availability Zones can contain one or more Host Aggregates and vice versa. When a VM is created, it is scheduled to one Availability Zone, and two Availability Zones cannot overlap with each other on nodes. However, Host Aggregates within an Availability Zone can overlap.\\\\
In Cinder, we do not have Host Aggregates. When a volume is created, it gets scheduled in one of the Availability Zones. Cinder Availability Zones also do not overlap with each other on nodes.

Kubernetes does not have these concepts like Regions, Availability Zones, Host Aggregates to keep things simpler. Kubernetes provides the same values through concepts like node labels, selectors, and resource-specific features like node affinity and anti-affinity in pods.

From the above observations, we see many concepts can be mapped while many cannot. It is not practical to discuss all such mappings here, and we leave this section with the above observations as a reference to build these mappings further as needed.

\subsection{KupenStack Reference Implementation}
Any KupenStack implementation should adhere to design and architectural principles specified in section 2.1. This section will cover suggestions and proposals for a reference implementation. The following proposed KupenStack implementation abstracts many OpenStack concepts with the intent of “OpenStack done the Kubernetes ways” and simplified OpenStack usage. This also depicts the wide range of possibilities that KupenStack implementations can have. We are proposing KupenStack for the first time, and we do not know what could be the best possible implementation for OpenStack as-Code custom resource definitions. Also, the proposed implementation will be available as open-source \cite{Kupensta29:online} for future research on KupenStack.

KupenStack deploys OpenStack services as containers and requires a high-level understanding of configuring and integrating these containerized services. KupenStack gives freedom to choose images for these containers from any source. For example, various open-source image projects like OpenStack-Ansible-LXC \cite{OpenStac10:online}, kolla \cite{GitHubop63:online}, LOCI \cite{GitHubop14:online} can be a source. Kubernetes has support for LXC containers and  OCI \cite{OpenCont50:online} containers both. The OpenStack community is also leveraging containers with OpenStack in many ways \cite{Containe68:online}.

KupenStack-controller can be implemented to deploy containers using custom manifest templates or using some existing open-source ones.  In this reference implementation, we propose to use some existing open-source OpenStack manifest templates from OpenStack-Helm \cite{GitHubop35:online}. Helm \cite{GitHubhe63:online} is a package manager for Kubernetes, and it neatly automates the deployment of Kubernetes manifest files. Helm gives lots of flexibility in customizing these manifests. This is why we propose OpenStack-Helm to be used along with our KupenStack reference implementation. Another benefit of adopting OpenStack-Helm is interoperability between multiple vendors. The OpenStack-Helm project can work as a standard reference for building any  KupenStack reference implementation compatible images. There are other automation projects which deploy containerized OpenStack, namely OpenStack-Ansible \cite{GitHubop72:online}, Kolla-Ansible \cite{GitHubop94:online}, Triple-O \cite{GitHubop29:online}, Kayobe \cite{GitHubop66:online} (it is Kolla-Ansible with bare-metal provisioning support using Bifrost \cite{GitHubop48:online}). Compared to the projects mentioned above, we consider OpenStack-helm as the best option for KupenStack. With OpenStack-Helm, we benefit from the automatic container lifecycle management feature of Kubernetes.

This Reference Implementation of KupenStack-controller implements operations like provisioning, maintaining, scaling, self-healing, upgrading, etc., over OpenStack-Helm.

\begin{figure}[h]
  \centering
  \includegraphics[width=6in, height=2in]{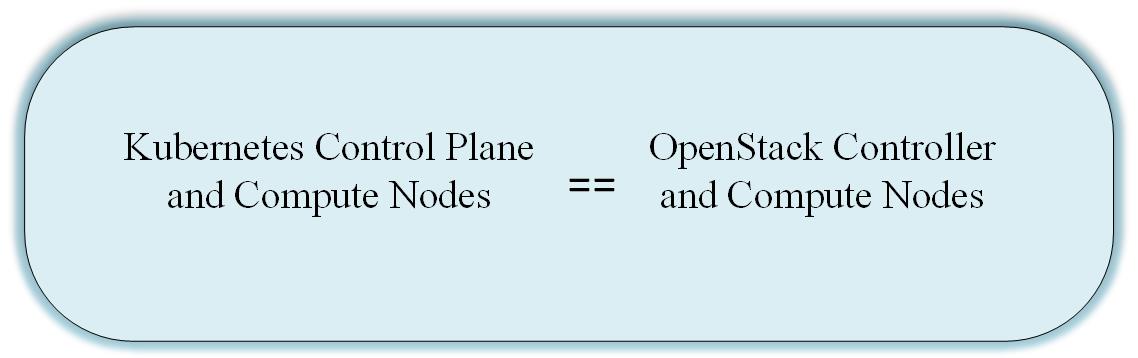}\\
  \caption{Node Mapping}\label{m}
\end{figure}

In this RI, we assume a one-to-one mapping between OpenStack and Kubernetes nodes for all resources and their placements. We keep OpenStack controller nodes to be physically the same as Kubernetes control-plane nodes, and OpenStack compute nodes to be the same as Kubernetes compute nodes as show in in Fig.\ref{m} . All custom resources (like Instances, Image, KeyPair, Router, Subnet) created by KupenStack-Controller are managed accordingly. KupenStack controller also performs intelligent cloud-native operations on OpenStack resources like self-healing of failed Instances similar to failed pods in Kubernetes.

\begin{figure}[h]
  \centering
  \includegraphics[width=6in, height=2in]{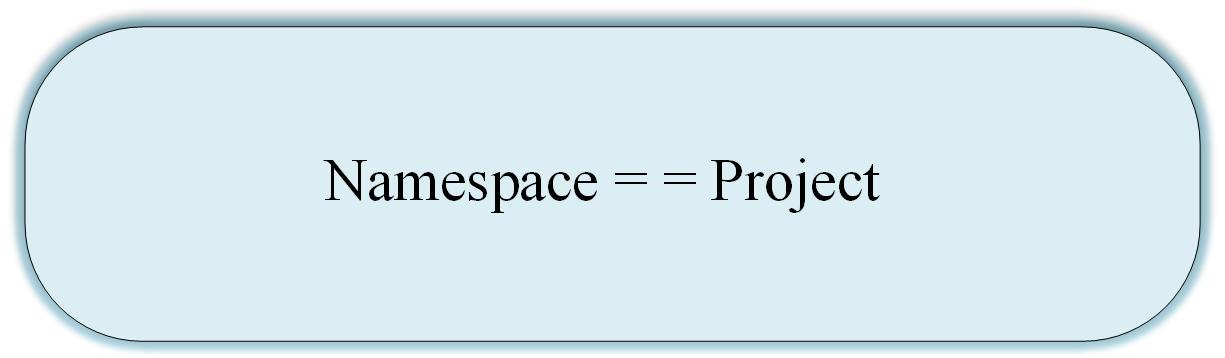}\\
  \caption{Logical Separation Concept}\label{mapi}
\end{figure}

For this implementation, we made Projects in OpenStack as abstract resources. Hence, rather than implementing separate custom resources for managing OpenStack Project, KupenStack maps them to Kubernetes Namespaces (using annotations in Kubernetes). As new Namespaces are created in Kubernetes, KupenStack creates corresponding Projects in OpenStack. For every KupenStack custom resource created in the Kubernetes namespace, it’s actual resource is created in the corresponding Project in OpenStack as shown in Fig.\ref{mapi}\\
Domains had nothing to map in Kubernetes. So, for our reference implementation, we will have only one Domain(i.e., default) in KupenStack because logically, Domain == Kubernetes cluster.

For authentication, there are three ways to integrate Kubernetes and OpenStack. However, in this reference implementation, we propose a different approach by ignoring OpenStack RBAC and authentication completely. We propose to use Kubernetes authentication only and depend on Kubernetes RBAC for all access management in KupenStack. Using Kubernetes RBAC and authentication,  OpenStack custom resources give unified and straightforward access management in KupenStack. Kubernetes has a flexible and widely adopted RBAC. This approach also allows us to change the default behaviour of Projects and resource visibility in OpenStack and builts OpenStack resources in the Kubernetes native way of logical separation through name+namespaces.

We are mapping OpenStack Region with the Kubernetes cluster (Region == Kubernetes Cluster) and have only one Region(named default).  Availability Zones implementation depends on individual OpenStack resources and services. For example, in Nova, we made a single Availability Zone(named default) for the Kubernetes cluster, i.e., Availability Zone == Kubernetes cluster. Virtual Machines created through custom resources can benefit from OpenStack Host Aggregates through labels and selectors design in Kubernetes. Hence, Host Aggregates are implemented as an abstract resource. This approach limits the use of OpenStack Regions, Availability Zone but gives perfect and smooth mapping to Kubernetes concepts. In this implementation, only Host Aggregates are used to build use cases like a failure domain as this is very similar to Kubernetes recommendations.

Similarly, more design choices can be made for OpenStack custom resources, and we will release more designs for other resources in the future. This section described the flexibility of making such choices for Reference Implementation.

\subsection{ Networking between Kubernetes Pods and OpenStack Virtual Machines}

It is desirable in a KupenStack environment that a Kubernetes Pod and OpenStack Virtual Machine are IP reachable to each other, and there can be other Kubernetes networking concepts (like service discovery, network policy) on them. There are open-source networking projects to achieve this with different approaches. 

Kuryr-Kubernetes \cite{GitHubop61:online} project tries to bring Neutron networking to Kubernetes pod in a Neutron-plugin agnostic way. Kuryr gives a cni plugin that attaches port to Kubernetes pod using Neutron plugin. When we look at the mapping between OpenStack and Kubernetes networking, we find the difference that Kubernetes leaves networking and follows a simple principle that all pods should be IP reachable to each other without NATing. However, OpenStack builds network infrastructure concepts like Network, Subnet, Router, etc., for its VM.
One approach for providing connectivity between pods and virtual machines can be to add extra interfaces to provide connectivity(while not disturbing the networking concepts of the two platforms). Multus-CNI \cite{GitHubk895:online} project is used to add an extra interface to the pod. Automation can be applied on top of Kuryr and Multus CNI through the controller, making all pods IP reachable to VM while not disturbing OpenStack networking. DANM \cite{GitHubno55:online} CNI is an alternative to Multus. The choice of meta-cni-plugin should not affect network performance as the principles on which these CNIs add multiple interfaces to pods is by delegating requests to other CNIs that add actual interfaces. The second approach could be to build network plugins for CNI and Neutron from scratch that understands both OpenStack and Kubernetes and the Calico \cite{GitHubpr45:online} project is one such popular open-source project. OpenStack-Helm projects have charts for configuring calico policies.

\section{Considerations for possible KupenStack integrations and innovations}
This section highlights the possible future integrations of KupenStack with various technologies and paradigms.

\subsection{Virtual Machine Images with tagging}
Process container technology open-sourced by Docker brought immutable images with tagging. Tagged images are handy in the GitOps world as this provides version control over image releases. In the process of making OpenStack cloud-native, we encourage to have similar tagging for Virtual Machines artifacts in KupenStack. Possibly, the OCI standard can be reused to store Virtual Machine artifacts. This gives benefits like completely reusing OCI image registries and hubs. KupenStack can internally implement a component that automatically converts these OCI artifacts to the respective image format for the hypervisor.  OCI standard is being used to create non-running container artifacts by open-source projects like Cloud-Native Application Bundles \cite{CNABClou8:online} and manifest-bundles \cite{enhancem37:online} ( Kubernetes SIG Cluster-Lifecycle \cite{communit19:online}) .

\subsection{Building Kubernetes concepts on top of KupenStack Resources}

With the introduction of KupenStack, there is space to build more Kubernetes concepts( like Daemonset, Deployment, RepicaSets, Jobs, etc.) on top of KupenStack custom resources for better management of OpenStack resources.

\subsection{Crossplane}
Crossplane is an open-source project that aims to build a unified control plane for multi-vendor resources running remotely. For example, a Crossplane user should be able to define a unified API through Crossplane Composite Resource Definition (XRD) \cite{Crosspla89:online} for SQL service offered from multiple vendors covering AWS, GCP, etc. Crossplane XRD assumes underlying Managed Resources to be implemented by the respective provider. As of this writing, Crossplane maintains the implementation of providers aws, gcp, alibaba, azure, and ibm in its open-source repos, but there is no implementation on managing remote OpenStack resources in any open-source project. These provider implementations in Crossplane only focus on provisioning and managing resources running remotely and keep no account of topology of workload placement as there is no cluster-level management. Crossplane assumes that cloud infrastructure is running remotely and doesn't focus on scaling, LCM, configuration management, and upgradation of the cloud infrastructure itself. It should be possible for Crossplane XRDs to integrate any implementation of KupenStack custom resources along with other public cloud alternatives.

\subsection{ Cloud Provider OpenStack}
Cloud-Provider-OpenStack \cite{GitHubku6:online} is another open-source CNCF project under SIG-Cloud-Provider \cite{communit65:online}. With the Cloud-Provider-OpenStack project, we can use OpenStack resources like Octavia-LoadBalancer and Cinder-Volumes as a provider to Kubernetes resources. For example, with the Cinder CSI Plugin in the Cloud-Provider-OpenStack project, we can get Persistence-Volume(PV) backed by Cinder inside Kubernetes. As of this writing, Cloud-Provider-OpenStack has some work on integration on Octavia, Cinder, Keystone, Manila, Barbican, Magnum components to Kubernetes. These integrations can be reused in KupenStack; this gives use cases like using Cinder volumes of OpenStack with Pods as well as VM.

\subsection{Cluster-API (CAPI) \& OpenStack-Magnum}
Cluster-API or CAPI \cite{GitHubku19:online} is another open-source CNCF project under SIG-Cluster-Lifecycle \cite{communit41:online}. CAPI gives a declarative way of provisioning, upgrading, and operating multiple Kubernetes clusters. CAPI can provision Kubernetes clusters on various targets like bare metal, AWS, GCP, Azure, and OpenStack. Integration of CAPI-OpenStack with KupenStack should be possible, giving us deployment topologies like Kubernetes-on-OpenStack-on-Kubernetes( KO3K). This can be used in a scenario where admins operating on KupenStack deployment want to provide a separate isolated Kubernetes cluster to their tenants. In some ways, this is multitenant Kubernetes. It is to note that the OpenStack-Magnum project is also a Kubernetes-as-a-Service project but for OpenStack. Ironically, for KupenStack both can be used. Hence choosing out of CAPI and OpenStack-Magnum depends on the user requirements.

\subsection{Airship}
Airship is an open-source project that declaratively automates cloud provisioning. A cloud software may be OpenStack, Ceph, Kubernetes, etc. For OpenStack, Airship internally uses OpenStack-Helm project, and this is where it overlaps with KupenStack. But Airship does not provide OpenStack resources as custom resources in Kubernetes, and to achieve this, Airship can integrate, and provision KupenStack based OpenStack-Helm cloud. So, hierarchy changes slightly, and KupenStack can become the middle layer between Airship and OpenStack-Helm project.

\subsection{Anthos}
Anthos is a hybrid and multi-cloud platform leveraging Kubernetes; it lets us run containerized Kubernetes applications without any modification across multiple platforms, clusters, and locations being on-prem or public cloud. As KupenStack builds OpenStack resources as first-class citizens into Kubernetes, this means in the future, we can see a true hybrid cloud, multi-cloud deployments for OpenStack through Anthos with no changes.

\subsection{KubeFed}
KubeFed(Kubernetes Cluster Federation) \cite{GitHubku86:online} is another open-source CNCF project which is under SIG-multicluster \cite{communit52:online}. KubeFed is a multi-cluster level project which aims to coordinate and manage resources over multiple clusters. KubeFed is extremely useful in complex multi-cluster use cases such as deploying multi-geo applications and failure domains. As an integration to KupenStack, similar Federation implementation can be extended to KupenStack resources. Here we would like to point out our example reference implementation in this paper. We proposed ignoring concepts of Regions and Domain in KupenStack, as ignoring those simplifies design and such higher-level integration on top like KubeFed become simpler to achieve and use in practice.

\subsection{Cloud-Native Network Function Virtualization Infrastructure}
There has been a lot of innovation and development in cloud technologies since the introduction of NFV. Over the last two years, as of writing this paper, more and more SDOs(Standards Defining Organizations) like ETSI, CNTT \cite{CloudiNf79:online} have adopted cloud-native in their definitions. Which makes Kubernetes a more suitable choice as seen in standard references like RI2 (Reference Implementation 2) \cite{CNTTCNTT36:online} in CNTT. But still, the unsolved problem happened to be an NFVi platform of choice that is cloud-native and supports both Container as well as Virtual Machine Workloads. KupenStack can greatly solve this problem as it is Kubernetes based and brings OpenStack to Kubernetes, a complete IaaS platform. This gives cloud-native platform to network orchestrators like ONAP \cite{HomeONAP64:online} for running 5G implementations having cloud-native CNFs and VNFs.

\subsection{KubeVirt, Kata Containers, etc.}

KubeVirt \cite{KubeVirt79:online} is an open-source project under CNCF that adds virtual machines support in Kubernetes. Also, there are Kata \cite{KataCont83:online} containers, which are lightweight Virtual Machines or MicroVMs that support running as Pods in Kubernetes. These are open-source projects where KupenStack overlaps, and if requirements are only running Virtual Machines, then KupenStack is more than just Virtual Machines. KupenStack brings a complete OpenStack IaaS.

\subsection{Terraform-Provider-OpenStack}
Terraform is an open-source infrastructure as code automation tool that uses a declarative configuration language. Terraform-Provider-OpenStack \cite{terrafor84:online} has a terraform implementation for provisioning OpenStack resources like Virtual Machines, Images, Keypair, etc. This is where KupenStack overlaps with Terraform-Provider-OpenStack with the advantage that KupenStack is Kubernetes control-plane based and can closely work with core Kubernetes resources. As far as automation is concerned, Helm charts can easily provide automation to KupenStack resources.

\section{Conclusion}

In this paper KupenStack is proposed  with the vision of making Cloud Native OpenStack which can be used in various use cases e.g. Hybrid Academic cloud, Edge, Multi-cloud, 5G etc. KupenStack can also be used with many open source projects like Airship, StarlingX, Magma etc. and work as an interoperability layer among many projects. This is the first version of KupenStack in which we have mapped various OpenStack resources with Kubernetes and in future, we will map more. We are also releasing source code on \cite{Kupensta29:online}. We will continue our work on KupenStack and make our future work public through this repo and also continue our implementation with other projects like KubeFed, Airship, etc. 









\bibliographystyle{unsrt}
\bibliography{twitref}
%


%








\end{document}